\documentclass[a4paper,fleqn,usenatbib, letters]{mnras}

\usepackage[dvipdfmx]{graphicx,xcolor}	
\usepackage{amsmath}	
\usepackage{amssymb}	
\usepackage{bm}           
\usepackage{threeparttable}
\usepackage{color}

\usepackage{newtxtext,newtxmath}

\usepackage[T1]{fontenc}
\usepackage{ae,aecompl}

\newcommand{\lsim}{\, \mbox{\raisebox{-1.ex}
{$\stackrel{\textstyle<}{\textstyle\sim}$}}\,}
\newcommand{\vect}[1]{\!\!\!\mbox{ \,\boldmath $#1$}}

\title[ Binary Periastron Time Shift with KL Oscillation]{ Cumulative Shift of Periastron Time of Binary Pulsar with  Kozai-Lidov Oscillation}

\author[H. Suzuki et al.]{
Haruka Suzuki$^{1}$\thanks{E-mail: suzuki@heap.phys.waseda.ac.jp (HS)},
Priti Gupta$^{1}$,
Hirotada Okawa$^{2,3,5}$,
and Kei-ichi Maeda$^{4,5}$
\\
$^{1}$Graduate School of Advanced Science and Engineering, Waseda University, Shinjuku, Tokyo 169-8555, Japan\\
$^{2}$Yukawa Institute for Theoretical Physics, Kyoto University, Kyoto 606-8502, Japan\\
$^{3}$Research Institute for Science and Engineering, Waseda University, Tokyo 169-8555, Japan\\
$^{4}$Department of Physics, Waseda University, Shinjuku, Tokyo 169-8555, Japan\\
$^{5}$Waseda Institute for Advanced Study {\rm (WIAS)}, 1-6-1 Nishi Waseda, Shinjuku-ku, Tokyo 169-8050, JAPAN
}
\date{Accepted XXX. Received YYY; in original form ZZZ}

\pubyear{2019}

\begin{document}
\label{firstpage}
\pagerange{\pageref{firstpage}--\pageref{lastpage}}
\maketitle

\begin{abstract}
We study a hierarchical triple system with the Kozai-Lidov mechanism,
and analyse the cumulative shift of periastron time of  a binary pulsar 
by the emission of gravitational waves.
Time evolution of the osculating orbital elements of the triple system 
is calculated by directly integrating the 
first-order post-Newtonian equations of motion.
The Kozai-Lidov mechanism will bend the evolution curve of
 the cumulative shift   when the eccentricity becomes large.
We also investigate the parameter range of mass and semi-major 
axis of the third 
companion  with which 
the bending of the cumulative-shift curve could occur within 100 years.
\end{abstract}




\section{Introduction}

The PSR B1913+16 system (Hulse-Taylor binary) found in 1975,
 is one of the most famous pulsar binaries 
 \citep{Hulse75}.
This binary pulsar has a highly eccentric close orbit: the semi-major axis is about 0.013 au and
 the eccentricity is about 0.617 \citep{Taylor76}.
Because of these features, the orbital energy of the system is extracted by the gravitational wave (GW) emission and its orbital period is decreasing gradually.  
This period shift has been detected for over 30 years as the cumulative shift of periastron time 
 through the observation of the radio signal from the pulsar.
This period shift effect
 is explained quite well by  the GW emission from the binary in general relativity \citep{Weisberg05, Weisberg10}.
It is the first indirect evidence of the existence of GW.

We have so far found many binary pulsars (see e.g. \citet{Lorimer08}).
Near some of the observed binary pulsars, there may exist a third companion.
In fact, the PSR B1620-26 system \citep{Thorsett99} and the PSR J0337+1715 system \citep{Ransom14} 
are triplet systems.
If the GW emission affects the orbital evolution of such a triple system,
one may wonder what kinds of evidence could be found in the observation
and how to observe them.

In a hierarchical triple system, 
the inner binary sometimes shows quite different orbital motion from that of an isolated binary.
The Kozai-Lidov (KL) mechanism \citep{Kozai62, Lidov62},  which is 
one of the most important phenomena in a hierarchical triple system, is 
particularly  interesting.
The KL-mechanism occurs when the inner orbit is inclined enough from the outer orbit.
The main feature of KL-mechanism is the secular changes of the eccentricity of the inner orbit and the relative inclination.
Both values oscillate amongst each other as a seesaw, that is, when  the
eccentricity decreases, the inclination increases, and vice versa.
When the eccentricity is large, the tidal dissipation becomes important and 
 the GW emission is more efficient because the periastron comes closer.
As a result, 
the KL oscillation will play  important roles in various relativistic astrophysical phenomena,
for example:
the merger of black holes \citep{Blaes02, Miller02, Liu17}, 
the tidal disruptions of stars by supermassive black holes \citep{Ivanov05, Chen09, Chen11, Wegg11, Li15}, 
and the formation of hot Jupiters \citep{Naoz12, Petrovich15, Anderson16}
or ultra-short-period planets \citep{Oberst17}. 

In this letter, we study the cumulative shift of the periastron time by the GW emission 
in a hierarchichal triple system with the KL-mechanism and discuss what we will observe 
through the radio signal from the pulsar.
This is the first theoretical prediction to indirect observation of GW from a triple system. 
The paper is organized as follows:
We briefly summarize the important features of KL-mechanism in \S\ref{sec:Kozai}.
After we explain our approach in \S\ref{sec:method}, 
we describe our models and present 
the results and discussions  in \S\ref{sec:result}.
In our result, KL-effects gives the different evolution curve of the cumulative shift from that of an isolated binary.
The condition where such a phenomenon occurs is also discussed.
The conclusion follows  in \S\ref{sec:conclusion}.

\section{Hierarchical triple system and Kozai-Lidov mechanism}
\label{sec:Kozai}

In this paper, we treat the so-called hierarchical triple system.
It  is a three-body system characterized by the following features:
The distance between the first and second bodies is much shorter than the distance to 
the third body.  We also assume that 
 the effect of the third body is much smaller than the gravitational interaction between the first and second bodies. As a result, we can separate the three-body motion into the
two-body inner binary orbit and the outer companion's orbit.

In a two-body problem in Newtonian dynamics,
 the  elliptical orbit is described by six orbital elements;
the semi-major axis $a$, the eccentricity $e$, the inclination $i$, the argument of periastron $\omega$, the longitude of ascending node $\Omega$, and the mean anomaly $M$.
Although these elements  are constant in the  isolated two-body system,
in the hierarchical three-body system, the perturbations from the tertiary companion affect
 the binary motion and modify  the trajectory from that of the isolated one.
Such a trajectory is not closed in general, but
we can define it as an {\it osculating orbit} at each time,
whose trajectory is approximated by the elliptic orbit with the above six 
orbital elements determined by 
the  instantaneous position and velocity\citep{Murray00}.
As for the outer orbit, we 
pursue the centre of mass of
the inner binary rotating around the tertiary companion.
It can also be described as another osculating orbit.
Hence, we  introduce two osculating orbits, which are called as inner and outer orbits:
The masses of the inner binary are $m_1$ and $m_2$, while  
 the tertiary companion has the mass $m_3$.  
We use subscriptions 'in' and 'out' to show the elements of inner and outer orbits, respectively.

In the hierarchical three-body system with large relative inclination, 
an orbital resonance known as Kozai-Lidov (KL) oscillation may occur.
This mechanism, which was discovered by \citet{Kozai62} and \citet{Lidov62}, 
is characterized by the oscillation of the inner eccentricity $e_\mathrm{in}$ and the relative inclination $I$. 
The relative inclination $I$ is defined as the argument between the inner and outer orbital planes.
This oscillation occurs in secular time-scale under the conservation of  the energy and angular momentum. 
Here we will briefly explain some key features of this mechanism 
in Newtonian quadrupole approximation for a restricted triple system ($m_2\rightarrow 0$)
(see e.g. \citet{Shevchenko17}).
 In such a restricted system,
 the oscillation results in the secular exchange of  $e_\mathrm{in}$ and $I$ with the conserved value of $\Theta$, 
which is defined by 
	\begin{equation}
		\Theta = (1-e_\mathrm{in}^2) \cos^2 I
\,.
		\label{eq:theta}
	\end{equation}
This approximation also gives the criterion of KL-oscillation as
	\begin{equation}
	0 \le |\cos I  |
			\ \le \sqrt{\frac{3}{5}} ,
		\label{eq:kozaiC}
	\end{equation}
which is equivalent to $39.2315^\circ \lsim  I  
			\lsim \ 140.7685^\circ $. 
The KL-oscillation time-scale is evaluated as 
	\begin{equation}
		T_\mathrm{KL} \sim P_{\mathrm{in}}\frac{m_1+m_2}{m_3} \left( \frac{a_{\mathrm{out}}}{a_\mathrm{in}} \right)^3
					 \left( 1-{e_\mathrm{out}}^2 \right)^\frac{3}{2} ,
		\label{eq:KozaiT}
	\end{equation} 
where $P_{\mathrm{in}}$ is the orbital period of the inner orbit.
 Note that some authors have studied this mechanism 
 for non-restricted triple system as well, including the general relativistic (GR) effect \citep{Naoz13a, Naoz13b, Will14a, Will14b, Will17}.
 For example, \cite{Naoz13a} showed that $\Theta$ is no longer conserved but oscillates 
	   if $m_2$ is not so small, even in the quadrupole approximation.
	   We will compare our simulation results with the Newtonian formula (\ref{eq:KozaiT}) in \S\ref{sec:result}.


The GR effect  changes these KL-criterion as \citep{Anderson17}
\footnote{
			\citet{Will14a, Will14b} has pointed out the difficulty of
           the GR correction and claimed that we have to take into account the
			 ``cross terms" between
			the Newtonian perturbation and the post-Newtonian precession.
			In our model, since we integrate the equations of motion directly,
			 the  effects of the 
		cross terms are automatically included.	
			}
	\begin{equation}
 		0 \le |\cos I|  \ 
			\le \sqrt{{3\over 5}\left(
											1-\frac{4}{9}  \epsilon_\mathrm{GR}
															\right) } \, ,
		\label{eq:KozaiC_PN}
	\end{equation}
 where $\epsilon_\mathrm{GR}$ is the GR correction term, written as
		\begin{equation}
			\epsilon_\mathrm{GR} = 
				\frac{3G}{c^2}\frac{ (m_1+m_2)^2 }{m_3}\frac{a_\mathrm{out}^3 \sqrt{1-e_\mathrm{out}^2} }{ a_\mathrm{in}^4 \sqrt{1-e_\mathrm{in}^2} } \,.
		\end{equation}
 		This value is derived from the double-averaged post-Newtonian Hamiltonian of two-body relative motion
		(See e.g. \citet{Migaszewski11}. The non-averaged original Hamiltonian is  given in \cite{Richardson88}). 			
The KL-oscillation time-scale is also modified when the GR  effect is taken into account.

\section{Periastron Time Shift
											by Gravitaional Wave Emission } 
\label{sec:method}

We study the cumulative shift of the periastron time of the inner binary,
which is caused by the GW emission.
In particular, we focus on the systems whose inner orbit is initially 
 inclined enough, such that the KL-oscillation will occur.

In order to solve the three-body system, we employ
 the first-order post-Newtonian equations of motion, called the Einstein-Infeld-Hoffmann (EIH) equations (\cite{EIH38})
\footnote{This equation could be derived from the Lagrangian given by Lorentz and Droste (\cite{LD17}).}:
	\begin{eqnarray}
	&&\hskip -.5cm
	\frac{ \mathrm{d} \bm{v}_{k}}{\mathrm{d} t}
			=-G\sum_{n\neq k} m_{n}\frac{\bm{x}_{k} - \bm{x}_{n}}{|\bm{x}_{k} - \bm{x}_{n}|^{3}}
				 \Big[ 
						1-4 \frac{G}{c^2}\sum_{n'\neq k} \frac{m_{n'}}{|\bm{x}_{k} - \bm{x}_{n'}|}
				 \nonumber \\
            &&\hskip -.5cm~~
                  		-\frac{G}{c^2}\sum_{n'\neq n} \frac{m_{n'}}{|\bm{x}_{n} - \bm{x}_{n'}|} 
						\left \{
                             		   1-\frac{(\bm{x}_{k} - \bm{x}_{n}) \cdot (\bm{x}_{n} - \bm{x}_{n'})}
                                     {2|\bm{x}_{n} - \bm{x}_{n'}|^{2}} 
						\right \} 
				\nonumber \\
            &&\hskip -.5cm~~
						+\left( \frac{|\bm{v}_{k}|}{c} \right)^{2} + 2\left( \frac{|\bm{v}_{n}|}{c} \right)^{2} 
						-4 \frac{\bm{v}_{k} \cdot \bm{v}_{n}}{c^2}
                  		-\frac{3}{2} 
                  		\left\{ 
								   \frac{(\bm{x}_{k} - \bm{x}_{n})}{|\bm{x}_{k} - \bm{x}_{n}|} \cdot \frac{\bm{v}_{n}}{c} 
						\right\}^{2}   
                 \Big]  
				\nonumber \\
           &&\hskip -.5cm~~
                 -\frac{G}{c^2} \sum_{n \neq k} \frac{m_{n}(\bm{v}_{k}-\bm{v}_{n})}{|\bm{x}_{k} - \bm{x}_{n}|^{3}}
                		(\bm{x}_{k} - \bm{x}_{n}) \cdot (3\bm{v}_{n}-4\bm{v}_{k}) 
				\nonumber \\
           &&\hskip -.5cm~~
				-\frac{7}{2} \frac{G^{2}}{c^2} \sum_{n \neq k}\frac{m_{n}}{|\bm{x}_{k} - \bm{x}_{n}|}
                		\sum_{n'\neq n} \frac{m_{n'} (\bm{x}_{n} - \bm{x}_{n'})} {|\bm{x}_{n} - \bm{x}_{n'}|^{3}}\,, 
		\label{eq:EIH}
     \end{eqnarray} 
where  $m_k$, $\bm{v}_k$, $\bm{x}_k$ ($k=1,2$ and $3$)
 are the mass, velocity and position of the $k$-th component of the system, 
$G$ is the gravitational constant, and $c$ is the speed of light.  
Eq.~(\ref{eq:EIH}) has been numerically integrated by using 6-th order implicit Runge-Kutta method, whose coefficients are obtained from \citet{Butcher64}.
We remark that the back reaction of GW emission on the orbital evolution corresponds to the 2.5 order post-Newtonian terms, 
which are not included in our calculation. It is because the effect of back reaction is quite small.

In order to set up initial conditions, 
we convert initial orbital elements of inner and outer orbits
 into the variables $\vect{x}_k$ and $\vect{v}_k$ in Cartesian coordinates,
 with its origin in the centre of mass of whole system 
 and with $x$-$y$ plane on the initial outer orbital plane 
(See e.g. \citet{Murray00}).
We integrate the above EIH equations (\ref{eq:EIH}) numerically.
We then evaluate the osculating orbital elements at each step from the 
numerical data of positions and velocities of the triple system (see e.g. \citet{Murray00}). 
Since the inner orbit is not exactly an ellipse, 
the obtained osculating elements are oscillating with small amplitudes in the cycle of inner orbit. 
Hence, we will take an average of the osculating elements for each cycle
to extract the global orbital elements  of the inner binary.
We then obtain the average semi-major axes $\bar{a}_\mathrm{in}$, $\bar{a}_\mathrm{out}$ and eccentricities $\bar{e}_\mathrm{in}$,  $\bar{e}_\mathrm{out}$, 
which may give the effective values of the orbital elements. 
Those elements will evolve in secular time-scale because of the effect of the tertiary body.

The orbital energy of inner binary, if it is close enough, dissipates by the emission of the gravitational waves, which causes a periastron shift as follows:
As derived in \citet{Peters63}, the period change for each orbital cycle is 
	\begin{eqnarray}
		\dot{P}_\mathrm{in}
				&=& -\frac{192\pi}{5} \left( \frac{P_\mathrm{in}}{2\pi} \right)^{-\frac{5}{3}} 
					 	\frac{G^2m_1 m_2}{c^5} \left( G(m_1+m_2) \right)^{-\frac{1}{3}} 
						\nonumber \\
				&\times&		
						\frac{1}{\left( 1-\bar{e}_\mathrm{in}^2 \right)^\frac{7}{2}} 
						\left( 
							1+ \frac{73}{24}\bar{e}_\mathrm{in}^2 
							+ \frac{37}{96}\bar{e}_\mathrm{in}^4 
						\right) 
						, \label{eq:Pdot}
	\end{eqnarray}
where $P_\mathrm{in}$ is the orbital period of the inner binary given by
	\begin{equation}
		P_\mathrm{in} = 2\pi \sqrt{\frac{\bar{a}^3_\mathrm{in}}{G(m_1+m_2)}} .
		\label{eq:P_in(a)}
	\end{equation} 
When the energy dissipation is evaluated for one binary cycle, the orbital elements can be treated as constant 
because  the back reaction of energy dissipation is small enough in such a timescale.
Here we have used the effective averaged values $\bar{e}$ and $\bar{a}$ instead of 
 the osculating orbital elements, $e$ and $a$, 
since  those elements oscillate in the inner-orbital period 
as  mentioned above
and may not reflect the global orbital elements of the inner binary.
 Since $\bar{e}$ and $\bar{a}$ depend on time, $\dot{P}_\mathrm{in}$ also changes in time.

In order to see this period shift, 
it is convenient to observe the cumulative shift of periastron  time 
$\Delta_P$ defined by
	\begin{equation}
		\Delta_P(T_N) =  T_N - P_{\mathrm{in}, 0}N,
		\label{eq:def_Delta_P}
	\end{equation} 
where $T_N$ is the $N$-th periastron passage time and $P_{\mathrm{in}, 0}$ is 
the initial orbital period of the inner binary. 
Using the definition
	\begin{equation}
		N = \int^{T_N}_0 \frac{1}{P_\mathrm{in}(t)} \mathrm{d} t\,,
		\label{eq:def_N}
	\end{equation}
where $P_\mathrm{in}(t)$ is the binary period at time $t$, which changes in time 
by the GW emission as
	\begin{equation}
		P_\mathrm{in}(t) =  P_{\mathrm{in}, 0} + \int^{t}_0 \dot{P}_\mathrm{in}(t') \mathrm{d}t' 
		\,,
		\label{eq:P_in(P_dot)} 
	\end{equation}
we obtain the cumulative shift of periastron time $\Delta_P$ as
	\begin{eqnarray}
		\Delta_P(T_N) 
			&=&  T_N-\int^{T_N}_0 \mathrm{d}t \frac{P_{\mathrm{in}, 0}}{P_{\mathrm{in}, 0}+\int^{t}_0\dot{P}_\mathrm{in}(t') \mathrm{d}t' }
				    \nonumber \\
			&=&  \int^{T_N}_0 \mathrm{d}t 
					\frac{\int^{t}_0\dot{P}_\mathrm{in}(t') \mathrm{d}t'}{P_{\mathrm{in}, 0}+\int^{t}_0\dot{P}_\mathrm{in}(t') \mathrm{d}t'} 
					\,.
					\label{eq:derivation_Delta_P}
	\end{eqnarray}
Since the emission energy of gravitational waves is very small, we usually expect
	\begin{equation}
		| \int^{t}_0\dot{P}_\mathrm{in}(t') \mathrm{d}t' | \ll P_{\mathrm{in}, 0} .
		\label{eq:assumption_Pdot}
	\end{equation}
In fact,  for Hulse-Taylor binary pulsar \citep{Weisberg05}, since we have
	\begin{eqnarray}
		P_\mathrm{b} &=& 0.32299\ \mathrm{day} ,\\
		\dot{P}_\mathrm{b} &=& -2.4184 \times 10^{-12} \ \mathrm{s/s}
		\,,
	\end{eqnarray}
the condition (\ref{eq:assumption_Pdot}) is true if $t \ll 3.656 \times 10^8 \ \mathrm{yr}$.
Hence, when we are interested in the time-scale such that $T_N \ll 10^8 \ \mathrm{yr}$, 
we can approximate $\Delta_P$ as 
	\begin{equation}
		\Delta_P(T_N) \approx \frac{1}{P_{\mathrm{in}, 0}} \int^{T_N}_0 \mathrm{d}t \int^{t}_0 \mathrm{d}t' \dot{P}_\mathrm{in}(t') .
		\label{eq:Delta_P}
	\end{equation}
If we assume $\dot{P}_\mathrm{in}(t) = \dot{P}_{\mathrm{in},0} =$ constant, 
we obtain
	\begin{equation}
		\Delta_P(T_N) \approx \frac{\dot{P}_\mathrm{in,0}}{2P_{\mathrm{in},0}} T_N^2 ,
		\label{eq:approx_Delta_P}
	\end{equation}
which was used in \citet{Weisberg05}.
However, in a hierarchical triple system with the KL oscillation, 
 $\dot{P}_\mathrm{in}(t)$ depends on time. 
 Hence, we 
evaluate  $\Delta_P$ by  Eq.~(\ref{eq:Delta_P}) with Eq.~(\ref{eq:Pdot}).

	Our analysis can be applied to general three-body system with a binary pulsar as long as the condition (\ref{eq:assumption_Pdot}) is satisfied.
		Here we stress that the cumulative shift of periastron time could be observed 
	    through radio signal from the pulsar even for very small GW emission such that 
        the  back reaction of GW emission on the orbital elements is negligibly small just as the case of 
        the Hulse-Taylor binary.


\section{Results and Discussions}
\label{sec:result}
To show our numerical results about the periastron time shift of a binary in a hierarchical triple system,
we shall choose the PSR J1840-0643 as an example\footnote{The most famous  Hulse-Taylor binary pulsar shows the cumulative shift of the priastron time given by Eq. (\ref{eq:approx_Delta_P}). However, the presence of a third stellar-mass object within $a_\mathrm{out} < 100\, \mathrm{au}$ was ruled out \citep{Smarr76}.}.
	This binary pulsar was discovered in {\it Einstein@Home} project, whose detail
	and the orbital parameters of the binaries are given in \citet{Knispel13}.
	The Doppler shift effect caused by the acceleration due to third body
	gives a constraint for the parameters of an outer orbit, if it exists.
	The Doppler time-scale $\tau_{\rm D} \sim c a_\mathrm{out}^2 G^{-1} m_3^{-1} $ should be longer than 
	 the characteristic age of pulsar $\tau_\nu$ defined by
		$
		\tau_{\nu} = \nu/(2 |\dot{\nu} |),
		$
	where $\nu(t)$ is the spin frequency of the pulsar.
	It gives the upper limit for the mass  of a third body and its distance.
	The PSR J 1840-0643 system has the characteristic age $\tau_\nu = 2.56 \times 10^6 \mathrm{yrs}$, 
	so it seems that this system also has a strict constraint on the presence of a third body like 
	Hulse-Taylor binary.
	This characteristic age was, however,  evaluated in the topocentric frame with 
the ansatz of an isolated binary.
	In the barycentric frame, the spin period 	is  increasing, 
   which looks unphysical  
	 \citep{Knispel13}.
	We therefore assess that this system has not yet had a strict constraint for the presence of a third body.
	
	 Assuming that this system has a third body, we set up initial values of the inner binary of our model by
	 using the observed parameters of this binary system, and analyse 
	 the cumulative shift of the priastron time $\Delta_P$.
	Table~\ref{tab:ini} shows the initial condition of our model.
	\begin{table}
			\begin{tabular}{ccccccc}
				\hline
				   orbit    &     $a$[au]        
           		&  $e$  &           $i$[deg]          &  $\Omega$[deg]        &   $\omega$[deg]   &    $M$[deg]     \\
				\hline
        		          inner    &   2.17373  &   0  &  60 &   0  &   -  &  0    \\
    outer   &       20.0   &   0  &  0  &   0 &   -   &  20   \\
				\hline
	\end{tabular}
		\caption{Initial orbital elements of our three-body system with $ m_{1}=1.4 \mathrm{M}_\odot, 
					m_{2}=0.16 \mathrm{M}_\odot$ and $m_{3}=30 \mathrm{M}_{\odot}$. $a$, $e$,  $i$, $\Omega$, $\omega$, and  
     			    $M$ are the semi-major axis,eccentricity,  inclination, longitude of ascending node,
         			argument of periastron, and mean anomaly, respectively. 
         			Those are fixed by the observational data for the inner orbit, while assumed for the outer orbit.
 					The argument of periastron $\omega$ can be arbitrary because the eccentricity is zero. 
					We have used this freedom to fix the axis of the reference frame.}
			\label{tab:ini}
\end{table}					

\begin{figure}
		\centering
		\includegraphics[width=8.5cm, bb=50 50 410 302]{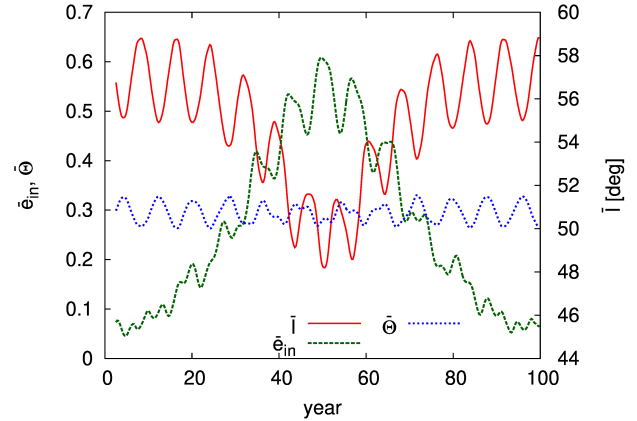}
		\caption{	
					The time evolution of the oscularing orbital elements  of our model;
					 the averaged eccentricity of the inner binary $\bar e_\mathrm{in}$, averaged relative inclination $\bar I$ 
					and averaged KL-conserved value $\bar \Theta$, for 100 yrs.
					The red, green and blue lines show the evolution of $\bar e_\mathrm{in}$, $\bar I$ and $\bar \Theta$, respectively.
					Many small oscillations are caused by the outer companion's motion, 
					while the KL-oscillation with time-scale of about 100 yrs is seen. 		
					}
		\label{fig:eccinc}
	\end{figure}

	\begin{figure}
		\centering
		\includegraphics[width=8.5cm]{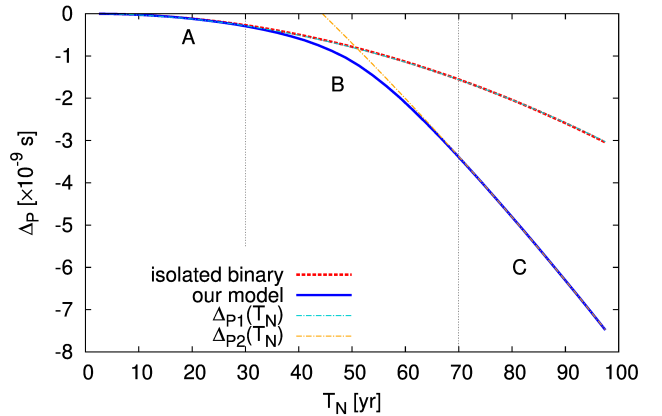}
		\caption{	
					The cumulative shift of periastron time $\Delta_P$ for 100 yrs.
					The blue line shows the result calculated for inner binary of our model, 
					which shows the deviation from  
					the red line named 'isolated binary' 
					 calculated for a simgle binary system.
					 In our model, the eccentricity is initially small, which $\Delta_P$ is fitted by the 
					 quadratic function $\Delta_{P1}(T_N)$ given in the text, but increases by the KL oscillation, 
					 which induces the deviation from the red dotted line, and eventually decreases again to small 
					 value, which makes the curve to another quadratic function $\Delta_{P2}(T_N)$ defined in the text. 		
					}
		\label{fig:CSPT}
	\end{figure}

The results for our triple-system model are shown in Figs.~\ref{fig:eccinc} and \ref{fig:CSPT}.
Fig.~\ref{fig:eccinc} shows the  time evolution of the averaged inner eccentricity $\bar e_\mathrm{in}$, 
averaged relative inclination $\bar I$ and averaged KL-conserved value $\bar \Theta$ for 100 yrs.
 $\bar \Theta$ is given by  Eq.~(\ref{eq:theta}) by use of $\bar e_\mathrm{in}$ and $\bar I$.
As for the evolutions of $\bar e_\mathrm{in}$ and $\bar I$, 
we find two kinds of oscillations with different time-scales: 
One is the period of 
outer orbit $P_\mathrm{out} = 15.92 ~\mathrm{yrs}$, and the other is 
 the secular oscillation time-scale $\sim 100 ~\mathrm{yrs}$, in which the effective eccentricity increases from 0 to about 0.6 
while the effective inclination decreases from $60^\circ$ to about $48^\circ$.  
The rapid oscillation with outer orbital period was also discussed in some papers \citep{Ivanov05, Katz12, Antognini14, Bode14}.
This secular oscillation of $\bar e_\mathrm{in}$ and $\bar I$ corresponds to the KL-oscillation.
Actually, the KL time-scale calculated by Eq.~(\ref{eq:KozaiT}), $T_\mathrm{KL} \sim 103.91~\mathrm{yrs}$, 
is consistent with the result of our simulation.
We remark that KL-conserved value $\bar \Theta$ is approximately conserved but shows small oscillation 
 with the period $P_\mathrm{out}$  around $\bar \Theta =0.3$. 
This is because our model is not an ideal 
 restricted triple system, that is,
$m_2$ is not a test particle \citep{Naoz13a} and 
the perturbation from outer body is not small enough to satisfy the condition for quadrupole approximation.
However, one important point is that the stable KL-oscillation is observed even in 
such a non-ideal 
restricted triple system.

Fig.~\ref{fig:CSPT} exhibits the time evolution of the cumulative shift of periastron time $\Delta_P$ 
for 100 yrs.
The result of our model is shown by the blue line.
As a reference, we also show the result of an isolated binary with the same initial data by 
the red dashed line, which is described by the quadratic function
	\begin{equation}
		\Delta_{P1} [{\rm s}]= -3.185\times 10^{-13}T_N[{\rm yrs}]^2\,.
	\end{equation}
The blue line of our model coincides with the red line initially (Period A in Fig.~\ref{fig:CSPT}),
 but  it deviates at $t \sim 30~\mathrm{yrs}$ 
and the discrepancy between these two lines becomes larger until $t\sim 60~\mathrm{yrs}$
(Period B in Fig.~\ref{fig:CSPT}).
This deviation of the blue line comes from 
the large amount of emission of gravitational waves, which is caused by 
the excitation of the eccentricity via the KL-mechanism. 
After $t\sim 70~\mathrm{yrs}$, the eccentricity decreases again, and then 
$\Delta_P$ is approximated by another quadratic function
	\begin{eqnarray}
		\Delta_{P2}[{\rm s}] &=& -3.185\times 10^{-13} \left( T_N[{\rm yrs}]+1.501 \times 10^2 \right) ^2
		\nonumber \\
		&&+1.206 \times 10^{-8}
		\,,
	\end{eqnarray}
whose curve is also given in the figure (Period C in Fig.~\ref{fig:CSPT}).
As a result, the curve of the cumulative shift will bend when the eccentricity becomes large by the KL ocsillation.
 
When we have a hierarchical triple system, as shown in Fig. \ref{fig:eccinc},
we may find an oscillation of the eccentricity via the KL oscillation by the long period observation of the orbital elements of a binary pulsar.
This has been already pointed out by several authors \citep{Gopakumar09, 	Portegies11}.
Here we have shown the cumulative shift curve will bend when the KL oscillation occurs.
It may be important because it will not only confirm
 the existence of a third companion but also  give the first indirect evidence of the GW
 from three-body system.

					Furthermore, this feature may be useful for real pulsar observation.
					In real observations, we sometimes cannot obtain
                 the observational data for some years due to some reasons; 
					 for example, we do not have the
					data of the Hulse-Taylor binary for a decade in 1990s
					because of major upgrades of Arecibo telescope \citep{Hulse94}.
					If this unseen region is completely overlapped with the period B, 
					it is impossible to recognize KL-oscillation only from orbital element data 
					because we miss high eccentricity state shown in Fig.~\ref{fig:eccinc}.
 			
However, with the plot of cumulative shift of periastron time like Fig.~\ref{fig:CSPT}, 
					we can conclude that 
 the bending in period B must exist from the observational data about the periatron time $T_N$ in both periods A and C.
					It is possible to judge whether KL-oscillation had happened or not by using our analysis
				    without the observational data in highly eccentric state.

One may be worried about the spin evolution of the pulsar 
because  the spin-orbit coupling appears in GR\citep{Barker75}, 
which may change the direction of the pulsar rotation axis. 
 If  the beaming direction of pulse signal is changed,
 the pulsar will disappear in the stage B.
However, 
following \cite{Liu17}, we find that the adiabaticity, which measures the ratio  of the KL oscillation time scale to 
the presession time scale, is very small in the present system.
It means that the spin of the inner binary evolves "non-adiabatically" 
even when the KL-oscillation occurs.
 As a result, the spin will be parallelly transported just as Newtonian case, and then 
the beaming direction will not change so much 
even in the KL-oscillation stage and the whole period (A, B and C) will be observed.

\begin{figure}
		\centering
		\includegraphics[width=8.5cm]{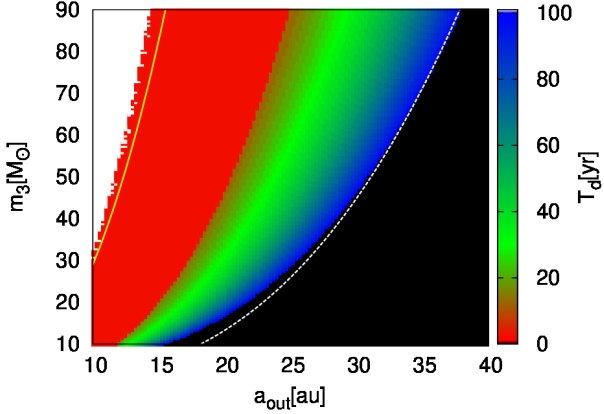}
		\caption{	
					The parameter range of third body with which the bending of cumulative-shift curve occurs within 100 yrs.
					The horizontal and vertical axes are the outer semi-major axis 
					and the mass of the tertiary companion, respectively.
					Color shows 
					$T_{\rm d}$, which is the minimum time such that the criterion of $	 (\Delta_P - \Delta_{P1})/ \Delta_{P1} > 1 $ is satisfied. 
					Black region means that KL-oscillation 
					does not occur in 100 yrs
					and the white dashed line corresponds to the theoretical boundary, $T_\mathrm{KL}=100~\mathrm{yrs}$.  
					In white region, the system became unstable; the yellow solid line is empirical criterion given in \citet{Blaes02}. 
					Detail explanation is in the text.
					}
		\label{fig:Detection}
	\end{figure}

Finally we shall discuss the possible parameter range 
 that the bending of the cumulative-shift curve occurs within our lifetime. 
In addition to the model given in Table~\ref{tab:ini}, we have performed our calculation for 19200 models 
	by changing the outer semi-major axis ($10\, \mathrm{au} \leqq a_\mathrm{out} \leqq 40\, \mathrm{au}$) 
	and the mass of the third body ($10 \mathrm{M}_\odot \leqq m_3 \leqq 90 \mathrm{M}_\odot$)
	and analysed in which  parameter ranges 
 the above bending will occur within 100 years. 
In order that such phenomenon occurs within 100 years, the KL time-scale should be less than 100 years.
So we investigate how many years are required 
 for the deviation of $\Delta_P$  from $\Delta_{P1}$ to become large enough.
For 19200 models with different $m_3$ and $a_\mathrm{out}$, we calculate $\Delta_P$
and  judge 
the deviation simply by the relative difference between $\Delta_P$ and $\Delta_{P1}$, i.e., 
	$	 (\Delta_P - \Delta_{P1})/ \Delta_{P1} > 1 $.
	We define $T_{\rm d}$ as the time when this criterion  is satisfied. 
Fig.~\ref{fig:Detection} shows the color contour  map of $T_{\rm d}$.
The black region in the bottom-right corner of Fig.~\ref{fig:Detection} corresponds to 
  $T_{\rm d}>100$ yrs.
The white dotted line shows the critical curve used by the Newtonian formula $T_\mathrm{KL} = 100 ~\mathrm{yrs}$ given by  Eq.~(\ref{eq:KozaiT}).
It is also found that the dependence of $T_{\rm d}$ on 
$a_\mathrm{out}$ and $m_3$ are consistent with
 the Newtonian formula $T_\mathrm{KL}$.  
					This is because the semi-major axis in the present model is large
 and 1PN effect is quite small.
The top-left white region in $a_\mathrm{out} < 15~\mathrm{au}$ and $m_3 > 30 \mathrm{M}_\odot$
 shows that the system becomes unstable if the initial parameters are in the region.
The yellow solid line is the empirical criterion for the instability given by \citet{Blaes02}:
	\begin{equation}
		\frac{ a_\mathrm{out} }{ a_\mathrm{in} } >\frac{2.8}{ 1-e_\mathrm{out} } 
															\left[ \left( 1+\frac{m_3}{m_1+m_2} \right) 
																	\frac{ 1+e_\mathrm{out} }{ (1-e_\mathrm{out})^\frac{1}{2} } \right]^\frac{2}{5}
																	\,.
		\label{eq:stabilityC}
	\end{equation}	
From Fig.~\ref{fig:Detection}, we find that 
		if this binary have the third companion with $m_3=10-100 \mathrm{M}_\odot$ and $a_{\rm out}\lsim 40 ~{\rm au}$,
		we may be able to see the bending of the cumulative-shift curve within 100 yrs through the observation of radio pulse.

\section{Conclusions}
\label{sec:conclusion}
We have  studied a hierarchical triple system with the Kozai-Lidov mechanism
and analysed 
 the cumulative shift of periastron time of a binary pulsar by the GW emission.
We have first proposed the theoretical calculation method of cumulative shift of periastron time 
					for general hierarchical three-body system with binary pulsar.		
Time evolution of the osculating orbital elements of the triple system 
is calculated by directly integrating the 
first-order post-Newtonian equations of motion.
We also investigate the parameter range of mass and semi-major 
axis of the third object  with which 
the above phenomenon could occur within 100 years.

For the inner binary of our triple-system model, 
we have employed the parameters of the binary pulsar 
 PSR J1840-0643 with zero eccentricity.
Assuming the existence of tertiary companion, for example, with the mass $30 \mathrm{M}_\odot$
and large relative inclination $I=60^\circ$,
we find that the Kozai-Lidov mechanism will bend the evolution curve of
 the cumulative shift when the eccentricity becomes large.
					Even if the data in period B is missed in realistic observation, 
					it is possible to judge whether the KL-oscillation had 
happened or not with the observed data in both periods A and C by using our analysis.
We have also investigated the parameter ranges of the outer companion around the binary, in
which the bending of the cumulative-shift curve due to KL-mechanism 
occurs within 100 yrs.
We find that 
this bending will occur in the range 
if a stellar or an intermediate mass black hole 
($m_3=10-100 \mathrm{M}_\odot$) exists at the distance within 40 au from the binary.

We are now performing our analysis for more general models because we will find 
many more triple systems in observations in near future.
Those results will be published elsewhere.
We are also interested in the GW emission from a triple system and its 
waveforms,  which study  is  in progress.

\section*{Acknowledgements}
We would like to thank Hideki Asada, Luc Blanchet, Naoki Seto and Kei Yamada for the useful information and comments. 
P.G.  is supported by Japanese Government (MEXT) Scholarship. 
This work was supported in part by JSPS KAKENHI Grant Numbers JP16K05362 (KM)  
and JP17H06359 (KM). 







%
%

\bsp	
\label{lastpage}
\end{document}